\documentclass[11pt]{article}
\usepackage[margin=0.75in]{geometry}
\usepackage[pdftex]{graphicx}
\usepackage{subfigure,amssymb}
\usepackage[margin=0.5in,textfont=small,labelfont=bf]{caption}
\usepackage{url}

\title{A computational model of cell polarization and motility coupling mechanics and biochemistry}
\author{Ben Vanderlei, James J. Feng, Leah Edelstein-Keshet}

\begin{document}

\maketitle

\abstract{
The motion of a eukaryotic cell presents a variety of interesting and challenging
problems from both a modeling and a computational perspective.  The processes span many spatial scales (from molecular to tissue) as well as
disparate time scales, with reaction kinetics on the order of seconds, and the deformation and motion of the cell occurring on the order of minutes.
The computational difficulty, even in 2D, resides in
the fact that the problem is inherently one of deforming, non-stationary domains, 
bounded by an elastic perimeter, inside of which there is redistribution of biochemical
signaling substances.
Here we report the results of a computational scheme using the immersed boundary method to address this problem. 
We adopt a simple reaction-diffusion system that represents an internal regulatory mechanism
controlling the polarization of a cell, and determining the strength of protrusion forces
at the front of its elastic perimeter. 
Using this computational scheme we are able to study the effect of protrusive and elastic forces on cell shapes on their own, the distribution of the
reaction-diffusion system in irregular domains on its own, and the coupled mechanical-chemical system.   
We find that this representation of cell
crawling can recover important aspects of the spontaneous polarization and motion of certain types of crawling cells. 
}

\section{Introduction}
Eukaryotic cell crawling is a complex process that involves interactions between mechanical forces and 
dynamics of biochemically active signaling
molecules. The deformation and motion of such cells are governed by a dynamic internal structure (the cytoskeleton) that is regulated by numerous kinds of proteins
and lipids. Unlike bacteria, whose motility is powered by flagella, eukaryotic cells move by a combination of protrusion, retraction, and
contraction. To crawl in a directed way, the cell has to first polarize and form a front, where the cytoskeleton assembles and leads to
protrusion, and a rear, where either contraction or passive elastic forces dominate. 
The determination of front and rear depends on
external stimuli (e.g. gradients of chemo-attractant) and has to be dynamic, sensitive, and yet robust. How cells manage this complex task is a
question of great interest in current cellular biology. 

Mathematical and computational researchers can provide techniques that help to dissect this complex process into simpler, more easily
understood modules. Prototypical ``in silico'' cells that share certain qualitative features with crawling cells allow us to address important
questions that are not as accessible in real cells. For example, in simulations, we can easily ask how mechanics decoupled from biochemistry (or vice
versa) affects the changing shape of a cell as it polarizes and crawls, and how the coupling between the two leads to emergent properties not present
in each on its own. We can explore the feedbacks between cell shape and the internal reaction-diffusion system that determines the chemical
distribution. Such questions are difficult to probe in experimental systems, and easy to test computationally. 

In order to investigate the mechanical-chemical coupling in cell shape and cell motion, we need suitable models and computational
algorithms. Simulating the motion of a eukaryotic cell is a difficult undertaking because it leads to solving chemical equations on a moving and deforming
domain. Such problems are recognized as numerically challenging.
Here we present the results of simulations that aim to meet this challenge at an intermediate level of complexity. We describe a
simulation package, developed from a composite of well-established techniques, that allows us to simulate the chemical polarization of the cell
(determining front and back), the forces of protrusion at the front (based on implicit growth of the 
cytoskeleton), and deformation of the cell, culminating in its motility in two spatial dimensions.  

We are not the first to simulate the motion of a eukaryotic cell.  In \cite{Gracheva04} a 1D continuum model is developed describing viscoelastic
properties of the cytoplasm in response to internal stresses generated by the contraction.  Treating the cell as a two-phase reactive fluid is a
well-studied idea in which the cytoplasm is treated locally as being a mixture, one phase of actin network and another of cytosol
\cite{Alt99,Herant03,Herant06}.  The immersed boundary method has been used to model explicitly the actin network and its adhesive links to a
substrate \cite{Bottino02}. The cell was modeled as a 2D elastic plate in \cite{Rubinstein05}.  Level set methods have also been used to include
chemistry dynamics on a deforming cell domain \cite{Iglesias,Wolgemuth2010}.  In most of these models, the asymmetries that lead to the motion of the cell are
imposed rather than self-organized.  Other models such as \cite{YuLiWang2008} couple chemical distribution on the cell edge with protrusion from a
central hub. A previous work \cite{Maree06} using the Potts model approach includes more detailed biochemistry, with  a Hamiltonian approach where
mechanical forces are implicit, rather than explicit. 

Our simulations belong to the class of mechanical/fluid-based models with the following features: (1) The perimeter of our cell is elastic. This
elasticity is attributable to the cortex, that part of the cytoskeleton directly adjacent to the cell membrane, generally composed of a network of
actin filaments. While we do not represent that network explicitly, we assign its elastic properties to the ``cell boundary''.  (2) We
model explicit forces, representing protrusion of the cytoskeleton, to the cell edge. Thus, we can also study the interplay between elastic and
protrusive forces. This differs from level set models (or cellular Potts models) that do not explicitly represent these mechanical properties of the
edge of the cell. 
Level set methods can prescribe a velocity, e.g.\ see \cite{Iglesias,Wolgemuth2010}, but the boundary curve has no mechanical properties of
its own. (3) Our platform is one of fluid-based computations. We can simulate both the diffusion and advection of substances inside the cell. In
Potts models and many other simulations, net flows of substances inside the cell are not tracked.  (4) Our simulation currently has a simple but
effective module that represents the internal self-organization of the cell.  Other models have included simple or more detailed internal biochemistry. For example,
Zajac et.\ al.\ studied the effect of the balanced inactivation model of \cite{levine2006} in their 2D level-set cell, and Mare\'e et.\ al.\
included three interacting regulatory proteins of the small GTPase family, and later added lipids such as PIP$_2$ PIP$_3$
etc \cite{Maree06}.  Keeping the regulatory system
simple but biologically faithful (to small GTPases) allows us to establish overall qualitative properties of the system, while pointing to the
emergent aspects of the coupling between chemistry and mechanics.  

In this paper we describe how the simulation package was assembled and present some its first results.
The mathematical model at the core of the simulation comprises two coupled systems of equations.  The first system is the mechanical model which
describes the fluid flow and the motion of the elastic cell edge.  The second system is a set of reaction-diffusion equations that describe a reduced
model for cell polarization analyzed in 1D in \cite{Mori08}.  These equations will be solved on the two-dimensional deforming domain that represents
the model cell.  The solution of the reaction-diffusion system will be directly coupled to the forces that the cell generates to crawl and deform. The
shape of the cell domain will, in turn, influence the solution of the the reaction-diffusion system. 

We use our platform to probe three distinct but interrelated regimes of behaviour. In the first, we consider the effect of mechanical forces that are
artificially prescribed, and investigate the shape of the cell for various force and elastic regimes. We show that even such simple cases lead to
shapes and motility relevant to some cell types. Next, we consider the regulatory reaction-diffusion system on a static but irregular domain. We show
how the shape of the domain influences the distribution of peaks of activity, and in particular, the effect of curvature on the ability of multiple
peaks to persist. Finally, we couple the mechanical and regulatory systems, assigning forces to the cell boundary in direct relation to the local level
of the active molecules. We show that such cells have very well-defined self-organized polarization and reasonable shapes, and that they move in a realistic manner.  

\section{Model Equations}

\subsection{Mechanical model}
\begin{figure}[ht!]
  \begin{center}
   \includegraphics[scale=1]{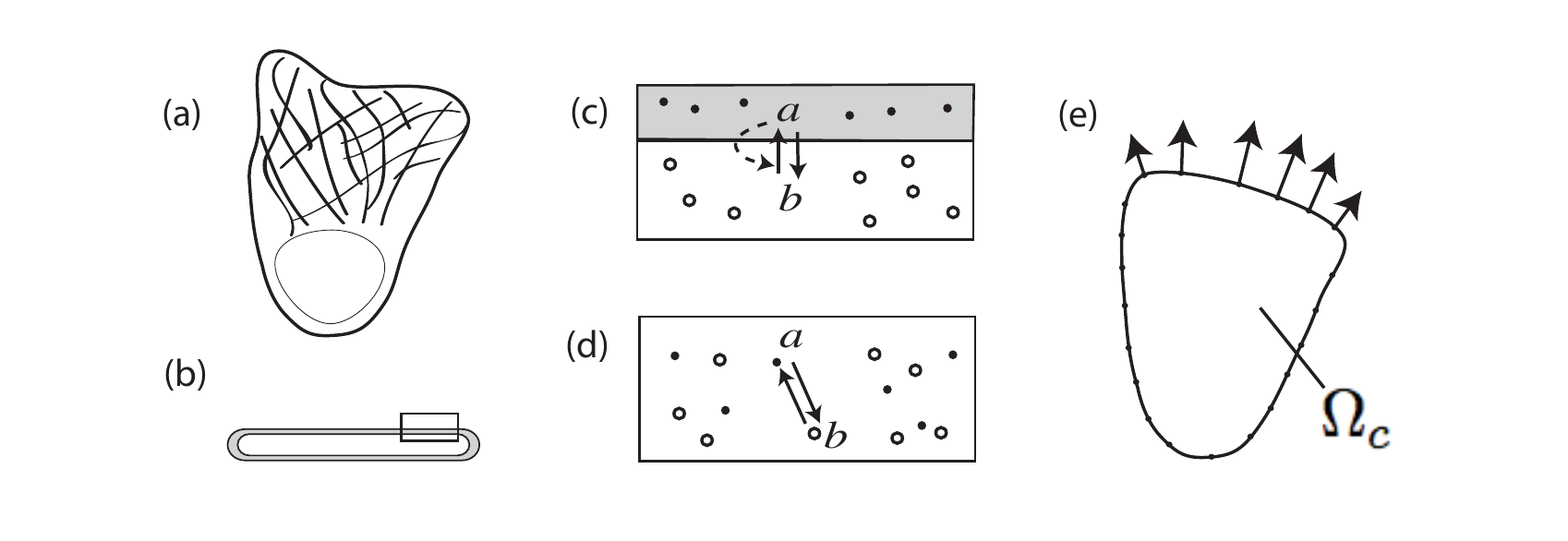}
  \caption{(a) Schematic diagram of a crawling cell in top down view (with nucleus and cytoskeleton). (b) Side view of a cell fragment with diameter
    $\approx 10\mu$m, and thickness $\approx 0.1-0.2 \mu$m showing membrane (grey) and cytosol (white). Rectangular region from (b) is enlarged in (c)
    to show the interconversion of two proteins, $a$ the active membrane-bound form and $b$ the inactive cytosolic form. The regulatory system
    includes (in)activation (with positive feedback, dashed arrow) and diffusion of $a, b$ everywhere inside the cell domain $\Omega_c$.  (d) In the
    simulation, we do not distinguish membrane from cytosol in the interior of $\Omega_c$, so $a, b$ occupy the same ``compartment'', with
    $D_b\gg D_a$. (e) We model the 2D top-down projection of the cell $\Omega_c$, devoid of nucleus and other structures, with elastic boundary
    representing the cortex, assuming a uniform thickness. The cell moves in the direction of the outward normal forces.  
 }
  \label{fig:schematic}
  \end{center}
\end{figure}

Our model is a two-dimensional (2D) representation of a cell, viewed from the top-down perspective. (See Fig.~\ref{fig:schematic}a,e).  We follow the
common convention of ignoring the cell body and nucleus as a passive load. This is reasonable in view of the fact that cell fragments (e.g. of keratocytes) devoid of nuclei can still migrate \cite{Verkhovsky1999}.
The cell is assumed to be resting on and adhering uniformly to a flat substrate. Typical dimensions of cell
fragments are 10-30 $\mu$m diameter, and 0.1-0.2 $\mu$m thickness. 

The regulatory module represents proteins (Rho GTPases) that have an active form, $a$,
bound to the cell membrane (grey surface in Fig.~\ref{fig:schematic}b,c), and 
an inactive form $b$, in the fluid cytosol (white portion of same panels). The interconversion $a \leftrightarrow b$ is regulated by other proteins
(GEFs, GAPs, not explicitly modeled) with positive feedback from $a$ assumed in $b \to a$ (dashed arrow in Fig.~\ref{fig:schematic}c). The membrane vs
cytosolic residence of the proteins affects their rates of diffusion $D_b\gg D_a$, but is not explicitly modeled, i.e. we abstract the view in
Fig.~\ref{fig:schematic} panel (c) by the simpler representation in (d).    

The cell is represented as a 2D domain $\Omega_c(t)$ (Fig.~\ref{fig:schematic}c). We associate a concentration of the signaling chemicals $a$ and $b$ 
to every element of area in $\Omega_c(t)$, since every such element corresponds to a small 
``sandwich'' of cytosol and membrane.
This description is in contrast with many current models \cite{Meinhardt1999,Ma2004,Krishnan2007,QingNie2009} that model chemicals
distributed only along the 1D cell boundary $\partial \Omega_c$.
The thinness of the ``cell'' is taken to mean that no significant gradients
form in a direction orthogonal to $\Omega_c(t)$.

Mechanically, the cell domain has an elastic perimeter enclosing a viscous incompressible fluid.  The elastic perimeter represents the cell cortex as
previously defined.
For the purposes of the mechanical description of the cell, it is only the perimeter of $\Omega_c$ that is endowed with elasticity and bears forces
due to the actin network.  So long as it is unstimulated, the model cell is symmetric, assuming a circular shape
with homogeneous internal chemical distribution; its perimeter is then under no elastic tension.  At this point in model development, the actin network and
the cell adhesion to the substrate are assumed implicitly, rather than modeled explicitly. That is, we connect the internal actin-regulating
biochemistry directly to forces of protrusion at the cell perimeter, rather than modeling the regulated growth of the actin network that actually
produces the force (but see \cite{Maree06,Rubinstein05,Rubinstein09}).

The mechanical model represents the interaction of the elastic membrane of the cell with a viscous incompressible
fluid.  To model this physical system, we use the well known formulation of the Immersed Boundary Method (IBM) \cite{Peskin02}.  
The key idea of this method is to replace the physical boundary
conditions at the cell edge with a suitable contribution to a force density term (\ref{force_density}) in the fluid equations
(\ref{momentum})-(\ref{continuity}), that are then solved by more conventional means.  To compute fluid flows, we use Stokes' equations. This utilizes
the well-known fact that at the 
cellular scale, the flow is at a very low Reynold's number, and inertial effects are negligible.
The other distinguishing feature of the IBM is the use of Lagrangian
marker points to track the boundary of $\Omega_c(t)$ for the purposes of computing a numerical solution.  The evolution of these material
points are then used to compute the elastic stresses in the membrane.

\begin{figure}[t!]
  \begin{center}
  \includegraphics[scale=1]{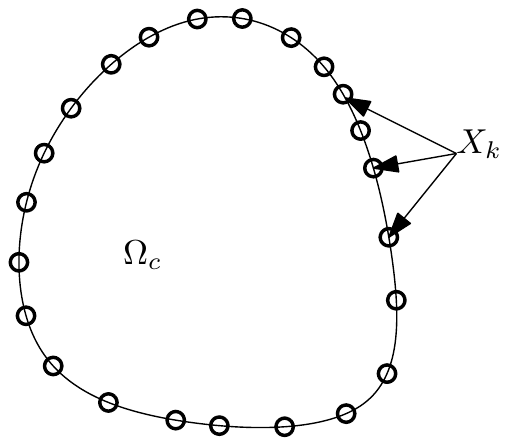}
  \caption{The cell domain $\Omega_c$ is the interior of a closed loop defined by $X_k$, a discrete collection of Lagrangian marker points.  The RD
    polarization model is defined only on $\Omega_c$, while the fluid equations are defined both inside and outside $\Omega_c$.}
  \end{center}
  \label{celldomain}
\end{figure}

The position of the membrane is given by a vector function ${\bf X}(s,t)$, where $s$ is arc 
length with respect to some reference configuration, and $t$ is time.  The boundary condition at ${\bf X}(s,t)$ is replaced by a
singular force density term in the fluid momentum equation.  In order to satisfy the no-slip boundary condition along the membrane, the boundary moves with
the local fluid velocity  ${\bf u}$ (\ref{membrane_motion}).  
Then the immersed boundary equations are

\begin{eqnarray}
 0 &=& -\nabla p + \mu \Delta {\bf u} +  {\bf f}(x,t), \label{momentum} \\
 0 &=&  \nabla \cdot {\bf u},   \label{continuity} \\
  {\bf f}(x,t) &= & \int_{\Gamma} {\bf F}(s,t) \delta(x-{\bf X}(s,t)) ds,  \label{force_density} \\
  \frac{\partial {\bf X}}{\partial t} &=& {\bf u}({\bf X}(s,t),t), \label{membrane_motion}
\end{eqnarray}
where $p$ is pressure, and $\mu$ viscosity.
The function ${\bf F}(s,t)$ is the magnitude of the singular force density (units: force per unit arc length $s$)
defined along the boundary, $\Gamma$, of the domain $\Omega_c(t)$.
It is composed of a force due to
the elasticity of the membrane, and a protrusive force due to the (implicit) polymerization of network,
represented by $h(a)$.  We take
\begin{equation}\label{force_line}
  {\bf F}(s,t) = F_{el} + F_{net},\quad F_{el} = \frac{\partial}{\partial s}\left[T(s,t){\bf \tau}(s,t)\right],
  \quad F_{net} = h(a){\bf n}(s,t).
\end{equation}
where $F_{el}$ is 
the elastic force, and $F_{net}$
is the protrusive network force.
In $F_{el}$, the quantity ${\bf \tau}(s,t)$ is a unit vector in the tangential direction, $T_0$ is the elastic modulus, and $T$ is the tension, assumed to take the form
\begin{equation}\label{tension_eqn}
  T(s,t) = T_0\left(\left|\left|\frac{\partial {\bf X}}{\partial s}\right|\right| -1 \right).
\end{equation}
In $F_{net}$, ${\bf n}(s,t)$ is the outward unit normal vector and $h(a)$ is a constitutive relationship between the local concentration of
the activated signaling system and the force generated by the network.  At this point, the mechanics equations are coupled to the biochemical reaction-diffusion
equations that depict the cell polarization and signaling.

\subsection{Biochemical model}

In \cite{Maree06} we considered a multi-layer signaling biochemistry that
regulates the cell polarization, and controls the growth and decay, protrusion and contraction of the actin cytoskeleton. Those models include such proteins as Rho GTPases (Cdc42, Rac, Rho), 
phosphoinositides (PIP, PIP$_2$, PIP$_3$), actin, and Arp2/3. However, for the purposes of establishing polarization and determining front and back of the cell, we later showed that a far
simpler model, consisting of a single Rho GTPase in active/inactive forms, suffices. That model, described in \cite{Mori08} in 1D, suits our purposes
here. It is sufficiently simple for the preliminary tests of our simulation platform, while producing results in 2D that have inherent features of
interest. 

Recall that the computational cell is a two-dimensional projection of a thin three-dimensional cell. 
The active form of the signaling protein, $a(x,t)$, and the inactive form $b(x,t)$,
diffuse in the domain $\Omega_c$ with disparate 
diffusion coefficients $D_a$ and $D_b$.  The two forms exchange at rate $g(a,b)$. 
The dynamics of our signaling model are thus governed by a pair of reaction-diffusion advection equations  
\begin{eqnarray}
  a_t + {\bf u} \cdot \nabla a &=& D_a\Delta a + g(a,b), \label{RDa},\\
  b_t + {\bf u} \cdot \nabla b &=& D_b\Delta b - g(a,b), \label{RDb}.
\end{eqnarray}
The advection terms in (\ref{RDa}, \ref{RDb}) account for the fact that the domain $\Omega_c(t)$ moves with respect to lab coordinates, and
carries the biochemistry along. 
For $g(a,b)$, we assume positive feedback enhancing the conversion of the inactive form $b$ to the active form $a$, but a constant rate $\delta$ for
converting $a$ to $b$:
\begin{equation}\label{RDc}
  g(a,b) = \left(k_0 + \frac{\gamma a^2}{K^2 + a^2}\right)b -\delta a.
\end{equation}
In this term $k_0$ is a basal rate of activation and $\gamma$ is the magnitude of the feedback activation rate.  The parameters $k_0$, $\gamma$, and
$\delta$ all have dimensions s$^{-1}$. The parameter $K$ has units of concentration of $a$ and we normalize concentrations so that $K = 1$.   

As shown in 1D in \cite{Mori08}, under appropriate conditions, the system (\ref{RDa}-\ref{RDc})
supports solutions in the form of a travelling wave that stalls inside the domain (The wave is then said to be ``pinned'').  In a pinned wave, the
domain is roughly subdivided into one region that supports a high plateau of $a$, while the remaining region has a low $a$ plateau; a sharp interface 
separates these zones, while the level of $b$ is relatively uniform throughout.  We refer to the high (low) levels of $a$ as $a^+$ (respectively
$a^-$).  Such a solution will be our description of a polarized cell, with $a^+$ the front portion and $a^-$ the rear portion of the cell.   
Two features of the model essential for this kind of polarized outcome are $D_b \gg D_a$ and
conservation of total $a$ and $b$. We therefore impose no flux conditions for both $a$ and $b$ at the boundary of $\Omega_c$.
A third necessary condition is that for a fixed $b=b_0$ within some range of values, the function
$g(a,b_0)$ has three steady-states ($a^-, a_h, a^+$), the outer two of which are stable i.e. that the well-mixed system is bistable in the variable $a$. 
We choose initial conditions that place $b$ within the appropriate range and $a$ close to the value $a^-(b)$. 

\subsection{Coupling biochemistry and mechanics}

We approximate the complex mechanical-biochemical coupling by assuming a direct link 
between the concentration of the activated signal protein $a$ and the local
force normal to the cell membrane. This assumption shortcuts the dynamics of growth of
the actin cytoskeleton and replaces the actin polymerization force by an ``effective
force of protrusion'' due to local activation of the Rho GTPase.
That is, when $a$ is above some threshold value $a_0$, we assume that there is
a local force directed outwards.
For our study we take the force magnitude $h(a)$ to be a piecewise quadratic
function, with adjustable parameters $H$ and $a_0$.
A convenient form for the relationship of force to the level $a$ of the activated protein is 
\begin{equation}\label{eqn:forcemag}
h(a) = \left\{ \begin{array}{ll}
 H\left( 1- \frac{(a-a^+)^2}{(a_0-a^+)^2}\right) &\mbox{ if $a> a_0$,} \\
  0 &\mbox{ otherwise.}
       \end{array} \right.
\end{equation}

This type of force distribution parallels experimentally observed distribution of actin filament ends that push on the leading edge in cells such as
keratocytes \cite{Keren08}. However, we stress that for now, our representation of the force distribution is meant to be qualitative. 

\section{Numerical Methods}

The primary difficulty in solving the model equations numerically is that the immersed boundary equations are coupled to the reaction-diffusion
system through the specification of the cell domain.  The solution of the immersed boundary equations determines the locations of
marker points, $X_k$, which comprise the discretization of the immersed boundary.  The location of these marker points define $\Omega_c$ on which the reaction
diffusion system is to be solved.  The solution of the reaction-diffusion system is then coupled to the immersed boundary model since it appears in
the forcing term $f(x,t)$.  To further add to the difficulty, the reaction kinetics happen on the time scale of seconds, much faster than the
motion of the cell (typically on the order of minutes or longer).  

We couple together several known and tested numerical methods and tools in the solution of our model equations.  We use the immersed boundary method for the formulation of
the mechanical boundary condition and discretization of the boundary \cite{Peskin02}, the method of regularized stokeslets for the flow computation \cite{Cortez01}, and the
immersed interface method for the solution of the reaction-diffusion system on the cell domain \cite{Jomaa08}.  In order to represent the interface and track its
location, we compute a level-set function.  In addition, we use adaptivity locally in time in order to make the computation robust.  

The solution of the model equations is carried out in the following steps:

\begin{enumerate}
\item {Compute the force distribution along the cell boundary due to membrane elasticity and protrusion.}
\item {Compute the flow field at the boundary marker points and on an internal cartesian grid that holds the signal concentrations.}
\item {Advect the membrane using the computed velocity.}
\item {Advect the solution of $a$ and $b$ according to the current fluid velocity.}
\item {Evolve the solution of $a$ and $b$ according to the reaction-diffusion system.}
\end{enumerate}

We define the three time steps $\Delta t_{fl}$,  $\Delta t_{ad}$, $\Delta t_{rd}$ for the
coupled system.  These represent the discretization of time in the fluid equations, the coupling advection term, and the reaction
diffusion system respectively.  The motivation for this method of splitting is that there is a wide seperation among the time scales of the different
processes and a time step restriction that is a result of the stiff immersed
boundary equations.  We allow for the possibility of taking multiple time steps in the fluid equations before updating the reaction-diffusion
system. A further reason for this split is that there is further computational overhead in the advection of $a$ and $b$.  This will be discussed shortly.
The time-marching scheme is thus comprised of the following steps. 
\begin{enumerate}
\item {Carry out $m$ time steps of the immersed boundary system.}
\item {Carry out a single time step of the advection of the signal concentration.}
\item {Carry out $n$ time steps of the reaction-diffusion system.}
\end{enumerate}
We will take $m\Delta t_{fl} = n\Delta t_{rd} = \Delta t_{ad}$ and set each of the time steps small enough to accurately capture the associated physical
phenomenon.  We discuss each of these steps in detail.

\subsection{Fluid velocity}
The simulation of the immersed elastic membrane is a well documented canonical problem that is treated by the immersed boundary method \cite{Newren07,Newren08,BAV:Stockie99,Tu92}.  We discuss here only the computation of the terms in (\ref{force_line}) that comprise $F(s,t)$
in our model.  
The restoring elastic force of the membrane,  $F_{el}$, is treated identically with other immersed boundary models.  
We take a centered difference discretization of derivatives comprising $F_{el}$, $T(s,t)$, and $\tau(s,t)$:
\begin{eqnarray}
  F_{el}(s_k,t) & = &\frac{T(X_{k-1/2},t)\tau(X_{k-1/2},t) - T(X_{k+1/2},t)\tau(X_{k+1/2},t)}{\Delta s}, \label{Fk_eqn} \\ 
  T(X_{k+1/2},t) & = & T_0 \left(\frac{|X_{k+1} - X_{k}|}{\Delta s} - 1\right), \label{Tk_eqn} \\ 
  \tau(X_{k+1/2},t) & = & \frac{X_{k+1} - X_{k}}{|X_{k+1} - X_{k}|}. \label{tauk_eqn}
\end{eqnarray}
Substituting (\ref{Tk_eqn}) and (\ref{tauk_eqn}) into (\ref{Fk_eqn}), one can rewrite the formula for $F_{el}$ as follows.

\begin{equation}\label{F_el_discr}
  F_{el}(s_k,t) = \sum_{i=k-1}^{k+1} \frac{T_0}{\Delta s}(|X_{i} - X_{k}| - \Delta s)\left(\frac{X_{i} - X_{k}}{|X_{i} - X_{k}|} \right)\frac{1}{\Delta s}.
\end{equation}

From (\ref{F_el_discr}) we see that the elastic force is equivalent to a force produced by discrete springs connecting the point $X_k$ with its neighbors.  The
resting length of the springs is $\Delta s$, taken in our calculation to be the spacing of the initial discretization of the immersed boundary
points.  We define the spring constant $ \sigma =\frac{T_0}{\Delta s}$.
Note that while $T_0$ has units of force, $\sigma$ has units of
force/length. 
To compute $F_{net}$, we determine the local value of $a$ using a piecewise constant reconstruction evaluated at $X_k$.  

The analytic representation of the Stokes flow that results from a single point force in the absence of
external boundaries is known as the Stokeslet. This velocity field is singular at the location of the point source. 
For the purpose of computing the Stokes flow that results from a discrete collection of point forces $F_k$ located at $X_k$, we use the method of
regularized Stokeslets \cite{Cortez01}.  The method 
replaces the singular force with a regularized force and computes an 
approximation of the true flow, which does not have the singularity.  We use the following regularization found in \cite{BAV:Tlupova09}.
\begin{equation}
  \label{eq:stokeslet}
  u(x) = -\frac{f_0}{4\pi \mu}\left[\frac12\ln{(r^2+\epsilon^2)} - \frac{\epsilon^2}{r^2+\epsilon^2} \right]   + \frac{f_0}{4\pi \mu}\left[f\cdot (x-x_0))\right](x-x_0)\left[\frac{1}{r^2+\epsilon^2} \right]
\end{equation}
Here, $f_0$ is the strength of the regularized singularity positioned at $x_0$, $r$ is the distance between $x$ and $x_0$, and $\epsilon$ is the regularization parameter.
Based on linearity of the Stokes' equations, we obtain the velocity due to a
collection of point forces by simple superposition.  (See \cite{BAV:Stokeslets3d} for an example of the application of the method to a
three-dimensional model of a swimming microorganism.)

In typical immersed boundary models, the solution of the fluid equations is found on a periodic rectangular domain with standard Stokes solvers.  
In some models, fixed boundaries are included by means of another discretized boundary, along which forces are distributed in order to produce zero
flow at the boundaries.  In this formulation, the problem is ill-posed when the integral of the forces does not vanish. 
Furthermore, the solution is only unique up to an arbitrary constant.  The work in \cite{Teran09} discusses this problem and presents a solution
method.

\begin{figure}[bt!]
  \begin{center}
  \includegraphics[scale=0.4]{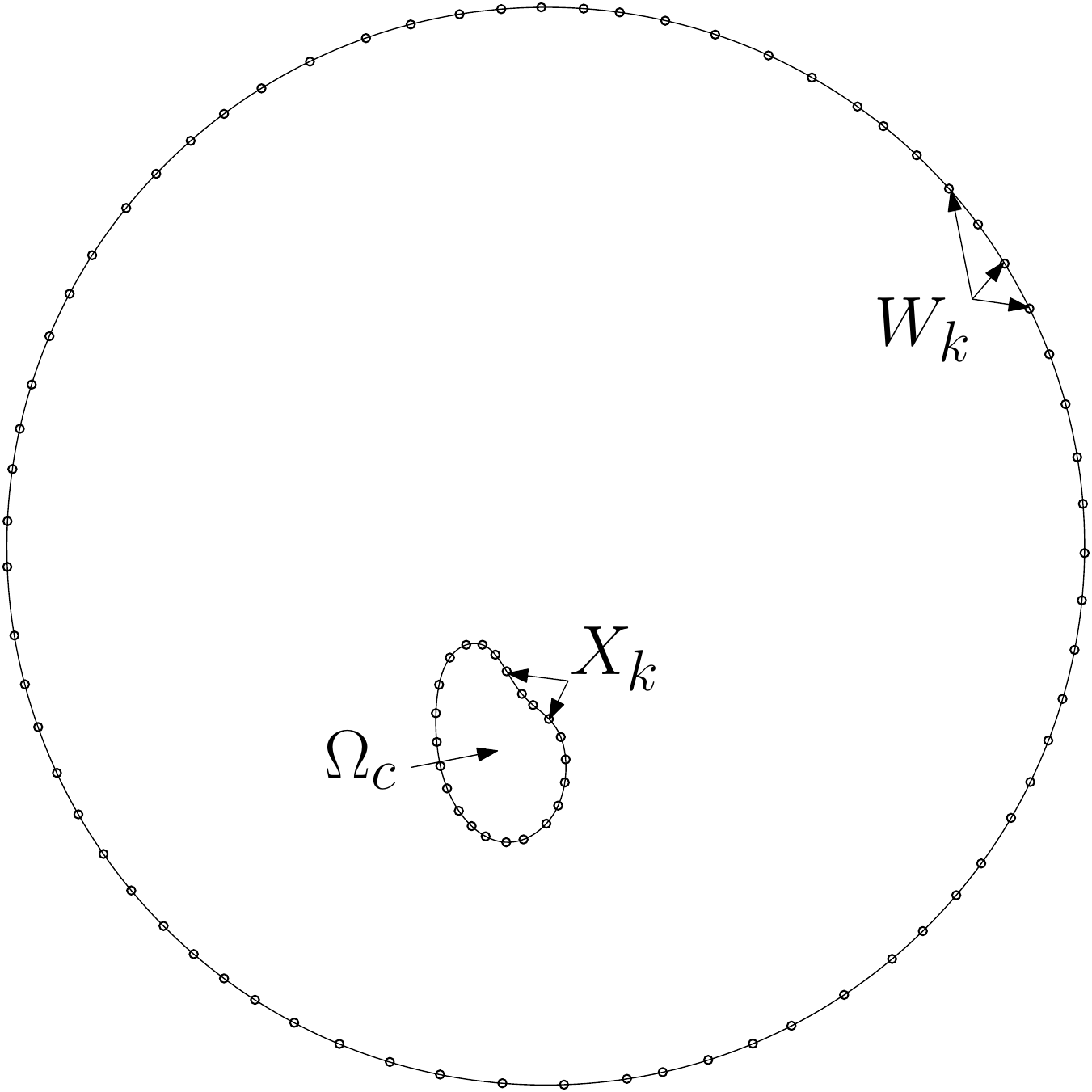}
  \caption{In order to enforce the condition of zero flow far away from the cell, additional immersed boundary points, $W_k$, are included in the
    computation as the representation of a fixed wall.}
  \label{walldomain}
  \end{center}
\end{figure}

A similar issue arises with the use of regularized Stokeslets due to the logarithmic growth of the Stokeslet as $r\to\infty$.  If the integral of the forces is non-zero, 
the velocity will be non-zero at infinity due to Stokes' paradox.  The inclusion of protrusive forces in our model means that we must take additional
measures to ensure that the flow will decrease to zero away from the cell domain $\Omega_c(t)$.  For this purpose, we include fixed walls in our computation, at
which we enforce the condition that the flow velocity vanishes.  This is achieved by discretizing the wall and distributing forces that cancel the
flow due to the logarithmic term.

The computation of the flow velocity due to a collection of $N$ point forces is the superposition of their individual contributions due to the
linearity of the equations.
The velocity at a point $X_j$ due to the point forces $F_k$ positioned at $X_k$ is thus a sum,
\begin{equation}
  \label{eq:stokessum}
  u(X_j) = \sum_k S_{jk}\cdot F_k.
\end{equation}
In this sum, the $S_{ij}$ are the terms in (\ref{eq:stokeslet}) that multiply the forces $F_k$.
Alternatively, this may be written as a matrix 
multiplication where $f$ is a vector of the forces, $u$ is a vector of the unknown velocities, and $S$ is a $2N\times 2N$ matrix,
\begin{equation}
 Sf = u.
\end{equation}
The inverse of this problem can also be considered, where the velocity at a collection of points is specified, and the forces that produce the
velocity are unknown.  The flow around a cylinder is demonstrated in \cite{Cortez01} using this method.

In our model, we include a collection of wall points $W_k$ as shown in Fig.~\ref{walldomain}, at which the velocity will be zero.  To compute the flow at a given point in the fluid  
domain due to the point forces $F_k$ at $X_k$, we take the following steps.

\begin{enumerate}
\item Compute the velocity at the wall points $W_k$.  Call this velocity $u_w$.
\item Solve a linear system for unknown forces $G_k$ positioned at $W_k$ that will produce $-u_w$ at $W_k$.
\item Sum over contributions to the velocity from $F_k$ and $G_k$.
\end{enumerate}

Although we must solve a dense system for $G_k$ at every time step, the matrix remains the same since the position of the wall points does not vary.
We can therefore compute a Choleseky decomposition once, and perform back substitution to obtain the solution needed at each time step.  We use the
standard routines in LAPACK for these calculations.

\subsection{Reaction-diffusion equations}
Next, we describe the solution of the reaction-diffusion system on the irregular cell domain.  We will address the motion the domain shortly, but
for now, we focus on approximating the solution of this system subject to no flux boundary conditions on a static domain of arbitrary shape.  The approach we
take is that of the immersed interface method.  The irregular domain is embedded in a rectangular domain on which a regular cartesian grid is used.
We produce a numerical solution at all grid points. For grid points outside the irregular domain, we set the numerical
solution to zero.  For grid points inside the domain, away from the irregular boundary, we use the standard centered finite difference
approximations to the second derivatives in the Laplacian.  
It remains to deal with grid points inside the domain next to the boundary, where the standard discretization fails.
These are denoted irregular points.  At these irregular points we must use a special 
finite difference formula that incorporates the no flux condition and the local geometry of the boundary.  To reiterate, the solution of the
reaction-diffusion system will be nonzero only on grid points determined to be inside the closed loop defined by the current location of the marker
points. 

\begin{figure}[ht!]
  \begin{center}
  \subfigure[acceptable]{\includegraphics[scale=0.7]{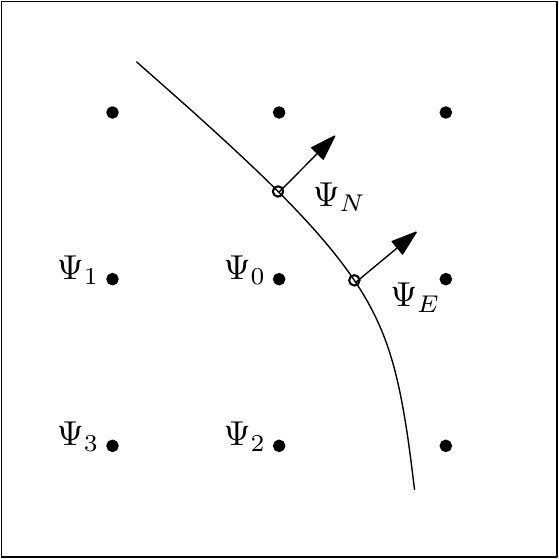}}\subfigure[acceptable]{\includegraphics[scale=0.7]{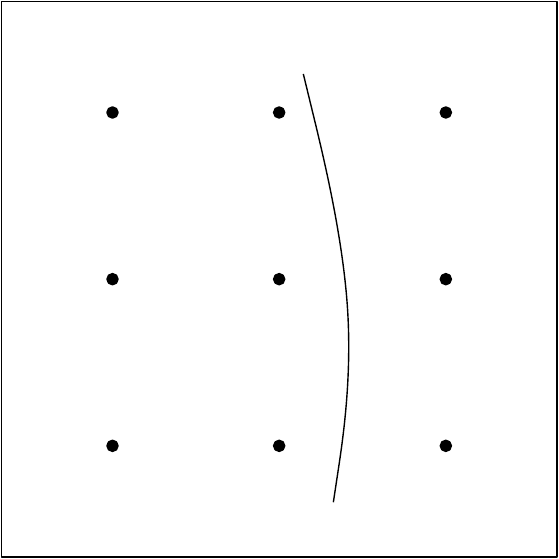}}\subfigure[unacceptable]{\includegraphics[scale=0.7]{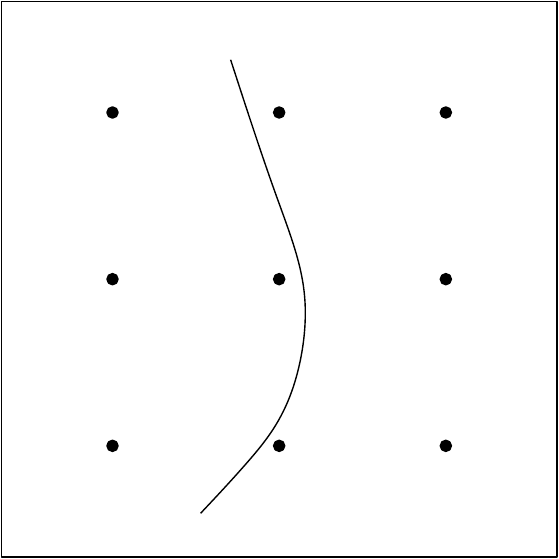}}\subfigure[unacceptable]{\includegraphics[scale=0.7]{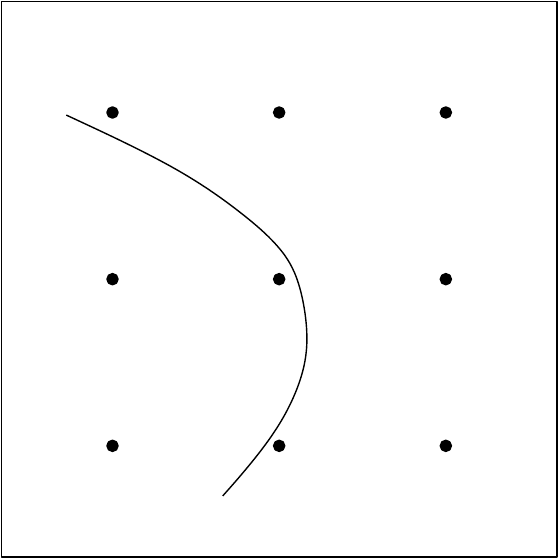}}  
  \caption{The points included in the finite difference scheme at an typical irregular point are shown in (a).
 $\Psi_N$ and $\Psi_E$ are the (unknown) values of the solution to the RD system on the ``cell perimeter''.  $\Psi_i$ for $i=0...3$ are the values
 of the solution at the grid points inside the computational domain $\Omega_c$.  The configuration in (b) is treated similarly.  At times when the
 boundary is in unacceptable configurations, we interpolate the solution to a finer mesh in order to proceed with the computation.}
  \label{irregularstencil}
  \end{center}
\end{figure}

We use the discretization scheme presented in \cite{Jomaa08} and refer the reader to that work for the details of the derivation.  We summarize here
the main idea and comment on the robustness with respect to geometry.  
See Fig.~\ref{irregularstencil} for the typical irregular point geometry.  If the boundary
conditions were Dirichlet, we would know the solution values $\Psi_N$ and $\Psi_E$ at the Lagrangian marker points. We could then write a discretization of $\Psi_{xx}$ using
$\Psi_E$ and the unknowns $\Psi_0$ and $\Psi_1$ and a discretization of $\Psi_{yy}$ using $\Psi_N$ and the unknowns $\Psi_0$ and $\Psi_2$.  However, for our model, 
we must employ Neumann (no flux) boundary conditions.
Thus, we do not have the values of $\Psi_N$ and $\Psi_E$.  Instead, these values must be interpolated from
$\Psi_0$,$\Psi_1$,$\Psi_3$, and $\Psi_4$ using the fact that $\Psi_n = 0$ along the boundary.  

A system of two algebraic equations can be written down for the unknowns $\Psi_N$ and $\Psi_E$.
We solve this system analytically and write down $\Psi_N$ and $\Psi_E$ explicitly in terms of $\Psi_0$,$\Psi_1$,$\Psi_3$, and $\Psi_4$.
In order to make the method robust, we enforce a lower bound on the determinant of the system.  This precludes cases where the
geometry is nearly degenerate as is the case when the interface is very near a grid point.  This is a common problem observed in similar numerical
methods.

The matrix of coefficients depends on the positions of the projections $\Psi_N$ and $\Psi_E$ as well as the unit normal vectors at these points.  We must
therefore have some continuous representation of the boundary constructed from the immersed boundary points $X_k$.  The representation we use is a
level set, as discussed in the next section.

A further complication we face in making the computation robust is that we must handle all potential configurations of the boundary with respect to
the grid.  For example, the discretization scheme just described will fail for configurations such as those indicated in Fig.~\ref{irregularstencil}c,d.  One approach would be
to write out discretizations for all configurations \cite{FogelsonKeener00}.  We choose instead to use an adaptive grid to avoid all cases except for the two
configurations shown in Fig.~\ref{irregularstencil}a,b (and their rotations). We perform refinements of
the grid and bilinear interpolations of the solution until the unwanted cases are eliminated.  The refinements are performed globally in space but
locally in time.  That is, for any time step we check for unwanted configurations.  If any are found, we refine the grid everywhere, interpolate the
solution to the fine grid, and repeat if necessary.  Once we have a satisfactory grid, we evolve the system and then project the solution back to the original grid. 

With the spatial discretization established, it remains to choose a suitable temporal discretization.  We use a fully implicit time method for the
diffusion operator in order to alleviate the otherwise prohibitive restriction on the time step size.  We use an explicit step for the reaction term.
This means that we must solve two sparse linear systems at every time step  
\begin{eqnarray*}
  A_a\,a^{n+1} &=& a^{n} + \Delta t \,g(a^n,b^n), \\
  A_b\,b^{n+1} &=& b^{n} - \Delta t \, g(a^n,b^n). 
\end{eqnarray*}
We use a GMRES method that is implemented in the library IML++.

\subsection{Advection and coupling}

In order to couple the mechanical system and the chemical system, we use a method first described in \cite{Tryggvason92} for the computation
of multi-fluid flows.  The idea is to construct a regularized version of an indicator function for the cell domain, that is, a function that is equal
to one in $\Omega_c$, zero outside $\Omega_c$, and varies smoothly within a transition region near the boundary.  Having such a function eliminates
the need to do complicated routines to track which of the cartesian grid points are inside $\Omega_c$ at any given time, knowing only the trajectories
of the Lagrangian markers $X_k$.  We can further use this indicator function to compute a level curve to use as a convenient representation of the
boundary of $\Omega_c$ when finding the discretizations needed for the irregular points in the Immersed Interface Method.

Given only the location of the marker points, we compute $I_{\epsilon}(x)$, the regularized indicator function which defines the domain on which the
reaction-diffusion system will be solved.  The gradient of the discontinuous indicator function $I(x)$ is zero 
everywhere except at the interface, where it has a singularity.  For the purpose of approximation, this singularity is represented again with the regularized
delta function $\delta_{\epsilon}$ and the gradient is then written as
\begin{equation}
  \nabla I_{\epsilon} = \sum_k \delta_{\epsilon}(x-X_k)n_k\Delta s_k.
\end{equation}
We then take the numerical divergence of $\nabla I_{\epsilon}$ to find $\Delta I_{\epsilon}$.  There only remains to solve the following Poisson problem in order to obtain
$I_{\epsilon}$.
\begin{equation}\label{indicatorproblem}
  \Delta I_{\epsilon} = \nabla \cdot G
\end{equation}
This problem will be solved on a rectangular domain which contains $\Omega_c$.  We position the rectangular domain such that its boundary is not near
$\partial \Omega_c$, so that the Dirichlet data for (\ref{indicatorproblem}) is zero.

Note here that a coarse discretization of the boundary of $\Omega_c$ relative to the cartesian grid will produce an $I_{\epsilon}$ with a level curve
that is oscillatory.  For this reason it is important that (\ref{indicatorproblem}) is only solved once on the coarsest grid.  This solution is then
interpolated to any refinements necessary for the solution of the reaction-diffusion problem.

\begin{figure}[t!]
  \begin{center}
  \includegraphics[scale=0.7]{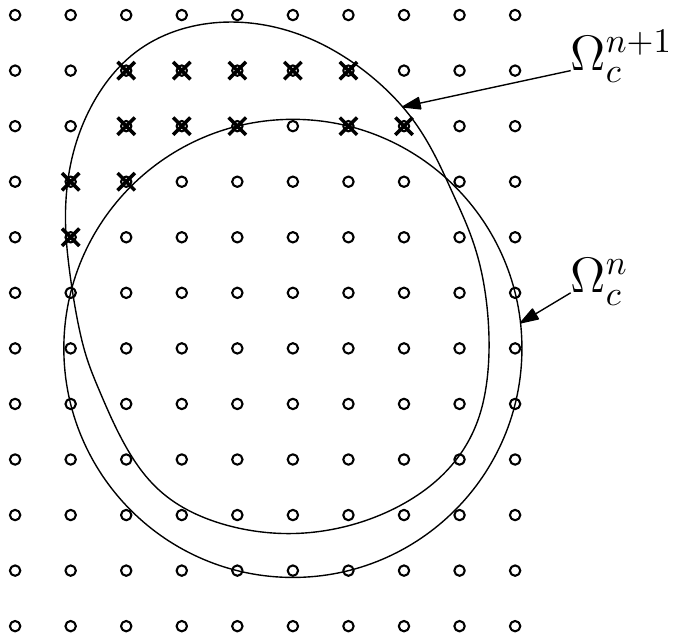}
  \caption{The known concentration at grid points in the domain $\Omega_c^n$ get advected with the fluid velocity to the grid points in the domain at
    the next time step $\Omega_c^{n+1}$.  As the cell perimeter moves, new grid points (marked with crosses) enter $\Omega_c$.  The advection
    defines values of the solution at these points in a way that is consistent with the model equations.}
  \label{advection}
  \end{center}
\end{figure}

The final detail that remains is to specify how to determine the solution of $a$ and $b$ at new points that enter $\Omega_c$ as it evolves with
time.  We need a way to map the solution of $a$ and $b$ from the previous domain to the current domain (Fig.~\ref{advection}).  We use an upwind 
discretization of the advection term that couples the reaction-diffusion equations to the flow equations.  Our Stokeslet solution to the fluid
equations allows us to compute the flow anywhere.  Specifically we can compute the velocity on the same grid as we are using for $a$ and $b$.    At
each grid point $x_{ij}$ in the current domain we use the fluid velocity to approximate the previous location of the material.

\begin{equation}
  x_{ij}^{n-1} = x_{ij}^n - \Delta t_{ad} u_{ij}^n.
\end{equation}
The location $x_{ij}^{n-1}$ will normally not lie on a grid point, but we can now use solution value of $a$ and $b$ at neighboring grid points to
interpolate their value at $x_{ij}^{n-1}$.  We have found that a simple piecewise constant interpolation is accurate enough to preserve the solution
profile through this mapping.

The advection scheme should conserve the total mass of the system $M_{tot}=\int_{\Omega_c} (a+b)\, dx$.  We approximate this
integral with a discrete sum $M_h$, again making use of the smooth indicator function:  
\begin{equation}
  \label{eq:totalmass}
  M_h = \sum_{(x_i,y_j)\in\Omega_c} I_{\epsilon}(x_i,y_j)(a(x_i,y_j) + b(x_i,y_j))h^2.
\end{equation}
In the sum, points that are near the boundary of $\Omega_c$ are weighted with a value less than $h^2$ since the area inside $\Omega_c$ that is
associated with boundary points is less than $h^2$.  
When preforming the advection we want to preserve the quantity $M_{tot}$ as well as the profile of the solution.  That is, we do not want to destroy
internal gradients through the advection step.  It is also important to keep the zero outward normal derivative that is consistent with the
discretization of the diffusion operator for boundary points.  For our application, we focus on preserving the solution profile and modify the
advection step so as to preserve $M_{tot}$ by adding or subtracting a small constant amount that may be lost in the upwinding scheme.  
We found that by so doing, we could retain the correct total mass to within 1\% of its fixed value at any given advection timestep. The mass change
due to the diffusion discretization is orders of magnitude smaller. Thus, by implementing the small correction, we find that the average total mass is
well-preserved. Experimenting with a variety of other advection schemes that preserve the mass, we found that other small oscillations in the solution
profile would tend to occur, leading to a magnified loss from the diffusion scheme.

\section{Results}

We present numerical experiments to illustrate separately the 
effects of protrusive forces along the domain edge $\partial \Omega_c$, the internal reaction-diffusion solver on a
static irregular domain $\Omega_c$, and the coupled system with self-organized forces emanating from 
the reaction-diffusion system.

\subsection{Mechanical model}

In our current simulations, the computational fluid is a means for distributing forces
and determining marker point motion. At present we use a fluid of similar viscosity inside and outside the domain.
Real cells have a much higher internal viscosity than their ambient environment, a feature that is not
yet represented.

We first consider a domain with no internal chemistry. Starting in each case with a
disk-shaped domain $\Omega(0)$, we prescribe half of the cell boundary to be the ``front'' and apply the protrusion force along this portion for the duration of the
simulation. We assign an outwards normal force whose 
magnitude has a parabolic profile that varies from a maximum of $H$ at the 
front center to zero at the cell sides.  

\begin{figure}[tp]
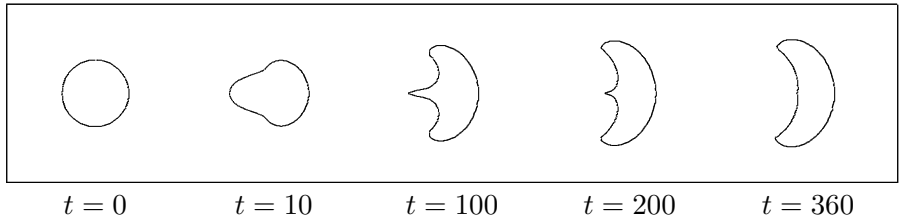

\begin{center}
  \include{mech_time_seq}
\caption{Mechanical model on its own. A time-sequence showing the initial configuration of the 
model cell, and its evolving shape over time. Simulation was
carried out using an elastic modulus $T_0=0.1$pN/$\mu$m, protrusion force $H= 12$pN/$\mu$m, and viscosity $\mu= 0.01$g/cm$\cdot$s.
By time $t=360$s, the cell is in steady-state motion at speed $v= 12\mu$m/s with the shape shown 
}
\label{fig:MechTimeSeq}
\end{center}
\end{figure}

\begin{figure}[h!]
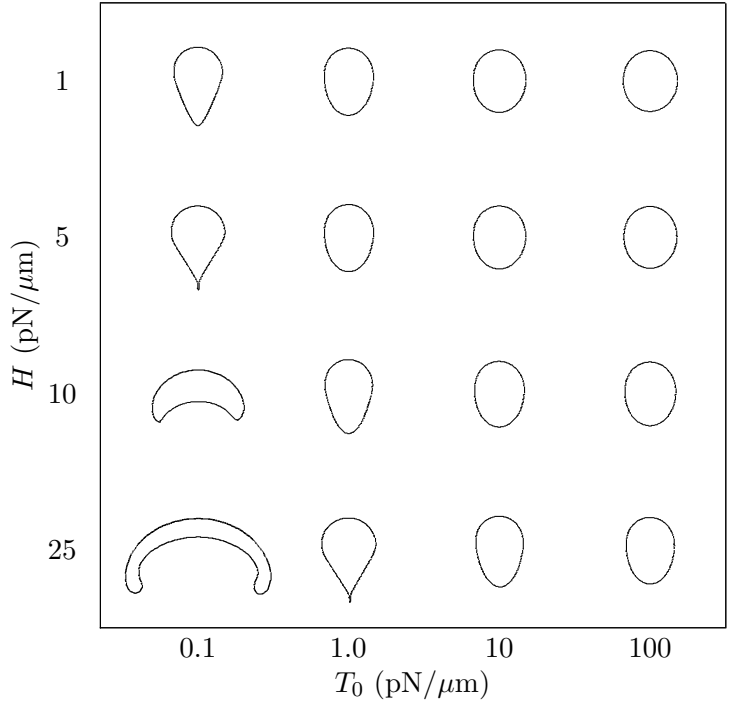

   \begin{center}
   \include{TvsH}
   \caption{Mechanical model on its own. A plot of cell shapes obtained as the protrusion force magnitude, $H$, and the elastic modulus $T_0$ vary,
     for a fixed, prescribed force distributed parabolically along the nodes that form the front half of the domain edge at time zero.  The direction of steady-state motion of these shapes is towards
     the top.  Viscosity was $0.01$g/cm$\cdot$s.  Note that the shape in the bottom left is a transient shape included only for completeness.  For
     force in the range of 1-25pN/$\mu$m, shapes obtained resemble a variety of biological cell phenotypes, from neutrophils to keratocytes.} 
   \label{fig:TvsH}
   \end{center}
 \end{figure}

As our curved domain edge is elastic, we can experiment with a variety of physical 
edge properties and examine how these affect the evolving shape of the domain. Note that 
this is one feature that makes the immersed boundary method distinct from level set methods, 
wherein the domain edge has no mechanical properties in and of itself.
The protrusive forces cause the perimeter of the domain to deform, and to move
in the direction of the force. The front edge that is under forcing tends to stretch as
it translocates, while the  rear edge (on which no force is prescribed) responds passively via
the elastic tension along the edge. 
Figs. \ref{fig:MechTimeSeq}-\ref{fig:TvsH}  illustrate solutions with typical shapes that are produced. 
Fig.~\ref{fig:MechTimeSeq} illustrates the transition in time between the initial circularly shaped cell and its steady-state
shape.  The transition in this example occurs over a time span of $360$s.  For cells with a smaller value of $H$, a steady-state shape is reached in less time.

To determine how mechanical parameters influence the shapes, and if
such mechanics-only solutions achieve a steady-state. To study this, we varied the three parameters in our mechanical system, the fluid viscosity $\mu$, the
membrane elasticity $T_0$, and the protrusion force $H$.  The ranges over which these parameters are varied include estimates for membrane elasticity,
and the forces and fluids relevant to them, since we are aiming to tailor our computation to solving problems on the scale of cell size and motion.  We
use the viscosity of water ($\mu = 0.01$g$\cdot$cm/s) since the fluid is most analogous to the cytosol, and typical forces (in pN) and edge tension
(pN per $\mu$m) estimated in \cite{Keren08}.   

Fig.~\ref{fig:TvsH} illustrates the variety of shapes obtained by varying the force on the leading edge, and the elasticity of the membrane
keeping the viscosity $0.01$g/cm$\cdot$s constant. For a relatively small value of $T_0\approx 0.1$pN/$\mu$m, we see a variety of cell shapes.  At low force magnitude, $H\approx 1$pN/$\mu$m, cell
shape resembles a teardrop (top left). At higher protrusive force, $H\approx 5$pN/$\mu$m, the teardrop 
develops a  longer tail, often resulting in cusp-like endpoint. At yet higher protrusion, the cell flattens out into a ``canoe'' shape, and then  a narrow crescent.  
Our current implementation of the IBM does not produce an accurate result when the magnitude of the protrusive force is
much greater than the force due to elasticity.  For example, when $T_0=0.1$pN/$\mu$m, values of $H$ greater than $12$pN/$\mu$m lead to large deformations 
that are not adequately resolved, and steady shapes then fail to form (bottom left is a transient shape, beyond the range of applicability of the method).  
 
When the elastic modulus is larger $T_0\approx 5-100$pN/$\mu$m, the transition  between shapes as $H$ is increased is less dramatic, since protrusion forces are counterbalanced by increasing
tendency to ``round-up''.  
Shapes produced for reasonable protrusion forces of 1-50pN/$\mu$m
and reasonable membrane elasticity ranges of 0.1-100pN/$\mu$m resemble a transition of shapes of cells such as neutrophils (teardrop) to
keratocytes (flattened canoes), cells whose motility is commonly studied experimentally.   

Varying the viscosity (of the computational fluid inside and outside the domain) has little effect on the   
shapes that develop. The dominant effect is to stretch the timescale over which those shapes are established. We ran tests for the range $0.01 \le \mu
\le 10$g/cm $\cdot$s and found similar shapes for corresponding values of the force and elasticity.

\subsection{Biochemical model}
\begin{figure}[b!]
 \begin{center}
\includegraphics[scale=0.5]{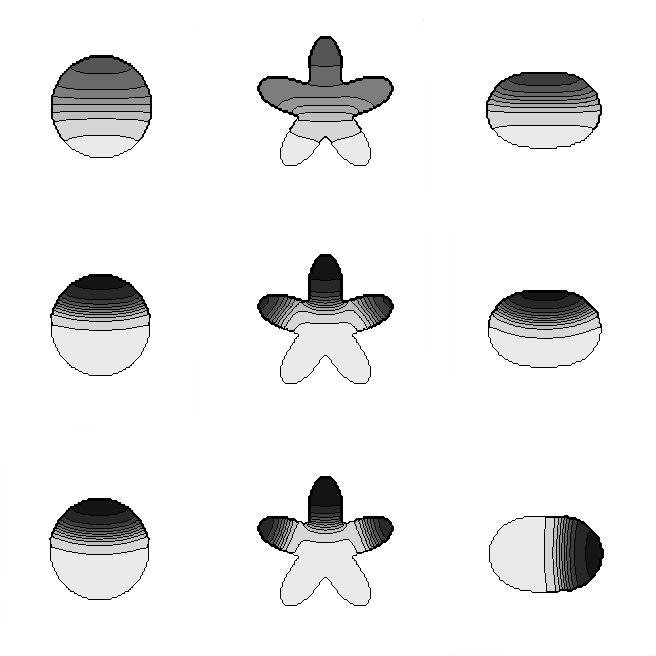}
\caption{The chemical model on its own. Solutions to the reaction-diffusion 
wave-pinning system given by Eqs. (\ref{RDa})-(\ref{RDc}) are shown in a variety of static geometries.
(a) circular disk (b) domain with convex and concave edges (c) ellipses. 
Time increases from top to bottom with snapshots shown at $t=20s, 50s$ and $200s$.
 The first row is the transient phase, during which there is a travelling wave.  The middle row shows the solution after the wave has been pinned.
 The bottom row shows the solutions after a long time.  The RD-solution correctly tracks the expected wave-like and stalled wave position, and shows
 the effect of concavity of the domain on the eventual location of its maximum. Reaction parameter values used were: $\gamma = \delta = 1$s$^{-1}$ and $k_0 =
 0.067$s$^{-1}$. Initial conditions were:  $a=0.27$ and $b=2.0$. A transient gradient term (\ref{eqn:stimulus}) was applied with $S=0.07$, $t_1=5$, and
 $t_2=15$.} 
\label{fig:rd_geomsWP}
 \end{center}
\end{figure}

We next tested the reaction-diffusion solver on its own. To do so, we selected a number of typical static domain shapes with various features and
solved a set of pattern-forming PDEs on the given domain. We considered a circular, star-shaped, and ellipsoidal ``cell'' shape. 
Shown in Fig.~\ref{fig:rd_geomsWP} are solutions to the ``wave-pinning'' system
Eqs.~\ref{RDa}-\ref{RDc} \cite{Mori08}.
The initial state of the system is spatially homogeneous with $a=0.27$ and $b=2.0$.  At $t=0$ a temporary
gradient is introduced via a transient reaction term $s(x,y,t) b(x,y,t)$ 
that is added to (\ref{RDc}), with 
\begin{equation}\label{eqn:stimulus}
s(x,y,t) =   \left\{ \begin{array}{rl}
S(1+x) & \mbox{for } 0 < t_1 \quad \quad \,\, \mbox{ and } -1 \le x \le 1 \\
S(1+x)\left(1-\frac{t-t_1}{t_2-t_1} \right) & \mbox{for } t_1 \le t < t_2 \,\,\mbox{ and } -1 \le x \le 1\\
0 & \mbox{for } t_2 \le t. \\
\end{array} \right.  
\end{equation}
where $S$ is a parameter that controls the magnitude of the stimulus.

  This gradient stimulates a portion of the cell to become activated,
meaning $a=a^+$ locally, and a travelling wave is initiated.  As discussed in \cite{Mori08}, as the level of $b$ is reduced by the reaction, the speed of the
wave slows and eventually stalls midway through the cell.
Based on related studies \cite{Mori08,Maree06} we anticipate solutions
to the reaction-diffusion system with relatively flat (low or high) plateaus, separated by a sharp interface. As shown in Fig.~\ref{fig:rd_geomsWP},
solutions have this character. In 2D with no flux conditions on the boundary of the cell domain, $\partial \Omega_c$, the level curves (shown in
Fig.~\ref{fig:rd_geomsWP}) also have to meet the curve $\partial \Omega_c$ orthogonally.  

The most interesting feature to observe is that the geometry of the domain influences the nature of the solutions. This is related to phenomena studied by
Mar\'ee et al (personal communication) showing the tendency of the kind of RD system considered here to minimize the curvature and the total length of
the interface. (Mar\'ee et al consider a related but more biochemically detailed wave-pinning system, but our simple caricature has similar
properties. See also \cite{Jilkine09}). In the star-shaped domain, three distinct peaks of activation can be maintained in three
``arms'' of the star, a phenomemon that would not occur in the simple circular domain.  We can understand this as heuristically as the energy of the
interface being at some local minimum.  
An application of similar ideas to the dynamics of plant Rho-GTPases in the pavement-cells of leaves has been proposed by V Grieneisen et
al. On the elliptical domain, in the short time-scale, the peak of activation is in the direction of the polarizing
stimulus. Once the stimulus is turned off, the polarization persists for a long time in this meta-stable configuration. Eventually however, the peak reorients to occupy the
high curvature pole of the ellipse: this allows for the interface length to shrink, and ultimately creates a stable configuration.

\subsection{Coupled mechanical-chemical model: self-organized cell shapes and motility}

Having tested the mechanical and chemical models separately, we now combine the two and allow the forces on the cell edge to be determined directly by
the solutions of the RD system. As described earlier, this leads to a number of new challenges, which are associated with solving a RD system on a moving
domain, that are accounted for using the solution algorithm described earlier. 
We used parameter values as in Fig.~\ref{fig:TvsH} for the mechanical properties, and values as in Fig.~\ref{fig:rd_geomsWP} for the RD system of
Eqs.~(\ref{RDa}-\ref{RDc}).  

\begin{figure}[b!]
  \begin{center}
  \includegraphics[scale=0.4]{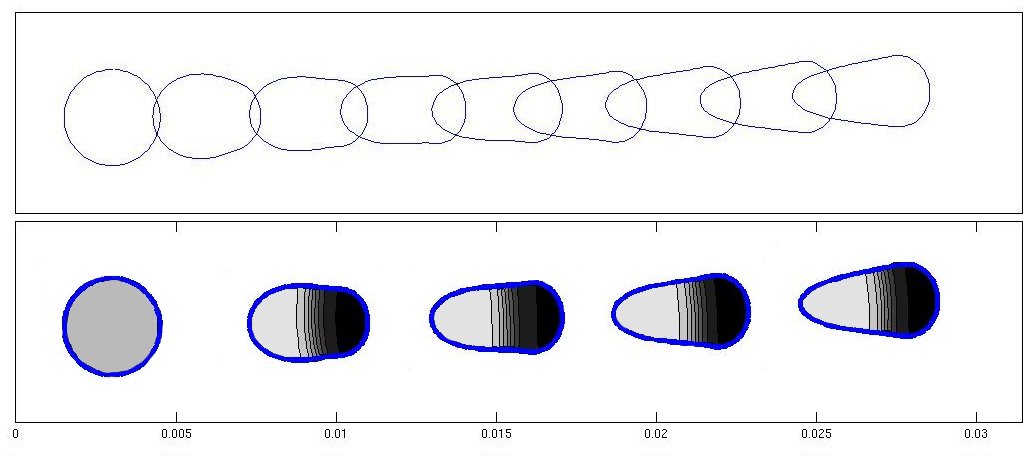}
   \caption{Mechanical and chemical model combined. Forces on the ``front'' edge are determined by the reaction-diffusion system inside the evolving,
     moving cell domain. Top:  A time sequence of cell shapes and positions, from $t=0 s$ to $t=800$s at intervals of $100$s starting from the circle
     at the left. The distance scale is in cm  ($10^{-4}$cm = 1 $\mu$m).   Bottom: Sample shots of the same moving cell at intervals of $200$s showing
     the internal chemical distribution. The cell is initially disk-shaped, with homogeneous internal concentrations of $a, b$. At $t=0$ a stimulus
     (biased towards the right) leads to chemical polarization.  RD equations (\ref{RDa}-\ref{RDc}) were solved in the domain
     $\Omega_c$ using parameter values as in Fig.~\ref{fig:rd_geomsWP}. The RD system very rapidly polarizes and
     maintains the polarization of the cell.  There is a slight change in the cell's direction of travel over long times.}
   \label{fig:WP_RDandMech}
  \end{center}
 \end{figure}

Initially, the system is in equilibrium both mechanically and chemically.
This means the membrane is relaxed to its equilibrium configuration and
the cell is disk-shaped. Both $b$ and $a$ are spatially uniform inside $\Omega_c$, and the active form is at a low steady-state level,
 $a(x,t)=a^-$.  At $t=0$, the same temporary gradient as was used in Fig.~\ref{fig:rd_geomsWP} is introduced via the transient reaction term in the signaling system.
The inactive protein, $b$ diffuses rapidly and is almost spatially uniform (but decreasing in level) as polarization proceeds.
As discussed in \cite{Mori08}, as the level of $b$ is reduced by the reaction, the speed of the
wave slows and eventually stalls midway through the cell. Here the forces are not prescribed; rather, they depend directly on the local magnitude of
the active protein $a(x,t)$ through (\ref{eqn:forcemag}).

Fig.~\ref{fig:WP_RDandMech} is a time sequence showing the evolution of the cell shape and internal chemistry.  The RD system
rapidly becomes polarized on a timescale of 30-60s and the cell starts to move and deform as a result. The teardrop shaped outline and constant speed
is then a steady-state moving solution to the combined RD-mechanical system.

\begin{figure}[t!]
  \begin{center}
 \includegraphics[scale=0.3]{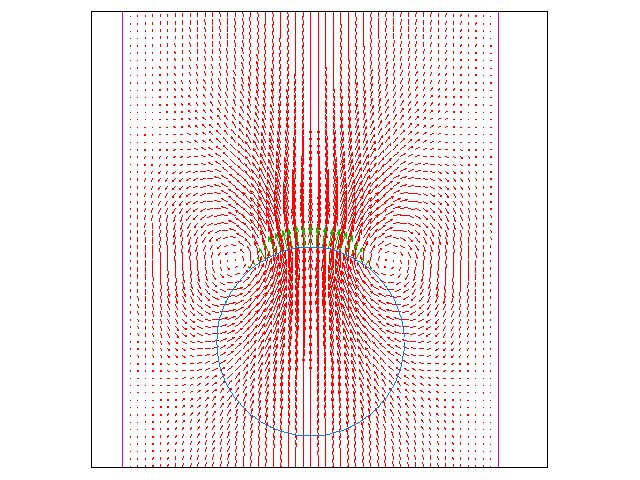}
   \includegraphics[scale=0.3]{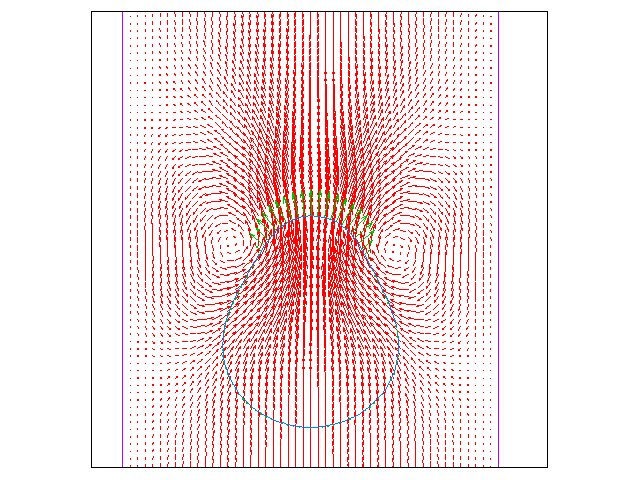}
   \caption{Computational fluid speed (red arrows) and chemically-regulated forces on
   the cell edge (green arrows) are shown at two times, close to the start of cell motion.
}
   \label{fig:FlowField}
  \end{center}
 \end{figure}

To illustrate the distribution of the fluid flow and pushing force, Fig.~\ref{fig:FlowField} shows two snapshots at the early stages of motion and
polarization.  The fluid velocity is highest at the front of the cell where the forces are largest.  Circulations in the flow that are directed from
the front of the cell around to the sides give rise to changes in the membrane shape.  After this transient regime, the flow becomes approximately constant in the
vicinity of the cell.  Since the membrane moves with the local fluid velocity, the steady cell shapes are observed.

\begin{figure}[ht!]
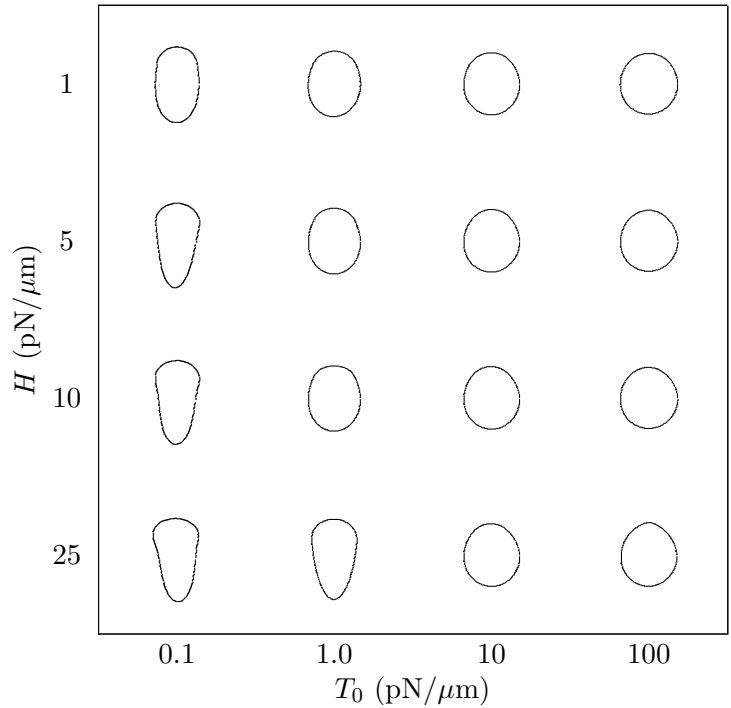

    \begin{center}
   \include{networkTvsH}
   \caption{Steady-state shapes predicted by the full mechanical-chemical system: As in Fig.~\ref{fig:TvsH} but with
  the leading edge regulated by the signaling system.  Cell shapes are much less sensitive to the magnitudes of $H$ in this self-organized
  system.} 
   \label{fig:networkTvsH}
    \end{center}
 \end{figure}

We asked how varying the mechanical parameters (elastic modulus $T_0$, viscosity $\mu$, and maximal force $H$) affect resulting shapes for the
coupled system.  Recall that $H$ sets the maximal level of force for $a=a^+$= maximal chemical
concentration. Fig.~\ref{fig:networkTvsH} shows the results, analogous to Fig.~\ref{fig:TvsH}. 
The chemical distribution within these ``cell'' outlines are similar to those of 
Fig.~\ref{fig:WP_RDandMech}, with rapid polarization on a timescale of $30-60$s and steady-state shapes by about $400$s.

Fig.~\ref{fig:networkTvsH} shows that increasing the maximal force magnitude in the chemistry-regulated cells tends to produce a transition from oval to
teardrop shape.
Fig.~\ref{fig:networkTvsH} similarly shows how varying the elastic modulus and the maximal force affects the behaviour. As expected, with larger
elastic modulus $T_0$, the steady-state shapes are rounder.   
A striking change seen from the mechanical to the mechanical-chemical system (Fig.~\ref{fig:TvsH} vs. Fig.~\ref{fig:networkTvsH}) is that the self-organized chemistry makes for much more stable and
persistent cell shapes. We can understand this as follows: with internal chemistry that is a self-organized polarized distribution, the nodes that are
assigned protrusion forces are governed by the fraction of nodes at high concentration of the reactant $a$. If the front stretches, nodes previously
experiencing large protrusive force leave the region of high $a$, and no longer carry such forces. This means that the self-regulated system is much
more stable, and resists forming wide fronts or narrow crescent shapes. 
 
\section{Discussion}

As described, the implementation of the immersed-boundary method leads to a convenient way to represent forces along the interface of the cell
boundary. Thus, we can capture actual forces and elastic deformations of the cell boundary (depicting the cytoskeletal cortex of the cell). This
allows us to track the cell shape deformation and the motion of the cell that results from  a coupling between internal reaction-diffusion and
boundary forces. Regions of highest activity (largest values of $a(x,t)$) develop spontaneously, determine the front, and result in distributed
forces. The linear elastic cell boundary reacting to the protrusion force, deforms to produce steady-state cell shape and motion. 

The simulations with mechanical forces alone, in the absence of internal regulation produced, a variety of shapes spanning morphologies of cells from neutrophils
(teardrop shapes) to keratocytes (canoe or crescent shapes) as protrusion forces increase relative to elastic forces. This suggests that the
properties of the cortex and the strength of actin-generated protrusive forces differ from one cell type to another, and at least broadly define the
cell's morphology.
Interestingly, once the simple regulatory system is incorporated, cell shapes are less varied and settle into a round/elliptical shape or teardrop, with
no crescents (compare Figs.~\ref{fig:TvsH} and \ref{fig:networkTvsH}). We can explain this fact based on the nature of the reaction-diffusion system,
(\ref{RDa}-\ref{RDc}), as previously discussed. The two key features of this system is that (1) it tends to form a sharp interface whose length tends
to be minimized by its dynamics (2) the interface meets the boundary orthogonally (by the no-flux condition). When forces tend to push nodes apart at
the cell front, the RD interface gets stretched. It reacts by finding a new configuration satisfying (1) and (2). This returns the peak of activity to
nodes that are not too far apart, and hence avoids the continual broadening of the front edge of the cell.  The interaction of the RD with direct
outwards forces on the boundary thus self-regulates the shape of the front, even as the force or membrane properties are varied. Needless to say,
other RD systems with interesting dynamics would, we predict, lead to other cell shape dynamics and steady-state shapes. 
  
In the current simulations, the IBM is used as a computational device to update
the cell edge, and to track the chemical distribution inside the cell domain. 
We do not interpret the internal fluid as cytoplasm, and the cell is not swimming. Indeed, a 
limitation of the current platform is the equal viscosity inside and outside the cell
that makes such interpretations flawed. Forces are present only
along the boundary curve, so that stress fibers of some cell types (e.g. fibroblasts) that lead to internal contractions are not yet
simulated. This could be done by including ``negative pressure'' terms in future fluid computations. The simulation is much more costly than a Potts
model, but comparable or less costly than finite element or level set methods. Part of the cost stems from the need for an
indicator function that tracks the cell inside and outside, and is recomputed at each time step. Such computational costs tend to limit the
exploration of parameter space or sets of assumptions that can be effectively explored.  

Applying the numerical scheme in its mechanical, chemical, and coupled variants, we found a
number of intriguing results. 
\begin{enumerate}
\item {Cell shapes generated by protrusion mechanics and passive elasticity, on their own, already lead to
morphologies that resemble some cell shapes.}
\item {Higher protrusion forces
(distributed along the front half of the cell) lead to flatter, more crescent-shaped cells, whereas larger elastic modulus leads to rounder cells
(Fig.~\ref{fig:TvsH}).}
\item {The wave-pinning mechanism of Mori et al \cite{Mori08} suffices to polarize the cell once a small bias is introduced
transiently. Previously, this system was studied in a 1D setting and in a rectangular domain \cite{Jilkine09}, and
we explore it here for the first time in irregular and deforming domains.}
\item {The shape of the domain feeds back on the chemical distribution. Parts
of the domain with higher curvature tend to ``attract'' the peaks of the active chemical (Fig.~\ref{fig:rd_geomsWP}.) This phenomenon was discussed by
AFM Mar\'ee in a related (but more biochemically detailed) cell biology context (personal communication).}
\item {Coupling the chemical and mechanical systems
leads to more stable cell shapes that are less sensitive to variations in such parameters as the magnitude of the protrusion force. }
\end{enumerate}
Here we concentrated on shapes produced exclusively by the wave-pinning biochemical model. These tend to form robust, broad plateaus of activity at
the ``cell front''.  Changing the internal regulatory module is a future step of interest, as it would reveal how qualitative aspects of the
regulatory system would affect qualitative aspects of cell shape.

These results can be compared with the recent work of Zajac and Wolgemuth \cite{Wolgemuth2010}. They effect cell deformation by modifying the spatial
variation in an adhesion coefficient.  The primary factor that determines our cell shapes is the strength of the polymerization force
relative to the membrane elasticity.  We also observed a dramatic difference in shape between cells in which the pushing force on the front was
imposed, and cells in which the front was regulated by the chemistry.   
In \cite{Wolgemuth2010} a velocity with a hyperbolic tangent profile is assumed.
Prescribing this velocity is most similar to our mechanical model without any chemistry, in which we 
prescribed a front and the profile of protrusive forces acting upon it.  In our coupled system we have gone one step further and produced a cell that
polarizes in response to a transient signal, and then maintains a steady shape and direction, without specifying the front of the cell in advance.    

Future development of such simulations will aim at addressing a number of questions. First, we will explore how contraction at the rear of the
cell affects cell shape (recall that here contraction is absent). We plan to investigate a variety of internal regulatory modules, including those
that have a greater diversity of spatio-temporal dynamics. As a later step, we plan to introduce a more explicit representation of the actin
cytoskeleton and of the lipids and proteins that regulate it. By gradually including such features one by one, we hope to learn how
tuning specific aspects of cell mechanics and biochemistry can lead to the repertoire of responses, deformation, and motility observed in real cells.   

\section{Acknowledgements}

LEK is funded by the Natural Sciences and Engineering Research Council (NSERC discovery and accelerator grants)
Canada. LEK is also funded by a subcontract from the National Institutes of Health (Grant Number R01 GM086882 to Anders Carlsson, Washington University, St Louis). 
JJF acknowledges support by the Petroleum
Research Fund, the Canada Research Chair program, NSERC (Discovery,
Accelerator and Strategic grants) and the Canadian Foundation for
Innovation. The computation was done at WestGrid.
We grateful to Lisa Fauci, Ricardo Cortez, AFM Mare\'e, V Grieneisen, A Jilkine, and A Mogilner for helpful discussions. 

\bibliographystyle{siam}
\bibliography{shape_study}

\end{document}